\documentclass[10pt,twocolumn,twoside]{IEEEtran}

\usepackage{amsmath,amssymb,epsfig,bbm}
\usepackage{graphicx,color}

\newtheorem{theorem}{Theorem}[section]

\newtheorem{remark}{Remark}

\def\eE{{\mathbb E}}
\def\pP{{\mathbb P}}

\def\rR{{\mathbb R}}
\def\cC{{\mathbb C}}
\def\zZ{{\mathbb Z}}

\def\proof{\noindent\textit{Proof:} }
\def\endproof{{\hfill $\clubsuit$ \medskip}}

\def\gen{\mbox{\footnotesize gen}}
\def\wAvg{W_{\mbox{\footnotesize avg}}}

\newcommand\cbrt[1]{{\sqrt[3]{#1}}}

\title{Bandlimited Field Sampling Using Mobile Sensors in the Absence
of Location Information}

\author{Animesh Kumar~\IEEEmembership{Member,~IEEE}
\thanks{A.~Kumar is with the Department of Electrical Engineering,
Indian Institute of Technology Bombay, India (email:
\texttt{animesh@ee.iitb.ac.in}).}}

\begin{document}

\maketitle

\begin{abstract}
Sampling of physical fields with mobile sensor is an emerging area.  In this
context, this work introduces and proposes solutions to a \textit{fundamental
question}: can a spatial field be estimated from samples taken at
\textit{unknown sampling locations}?

Unknown sampling location, sample quantization, unknown bandwidth of the field,
and presence of measurement-noise present difficulties in the process of field
estimation. In this work, except for quantization, the other three issues will
be tackled together in a mobile-sampling framework.  Spatially bandlimited
fields are considered.  It is assumed that \textit{measurement-noise affected}
field samples are collected on spatial locations obtained from an
\textit{unknown renewal process}.  That is, the samples are obtained on
locations obtained from a renewal process, but the sampling locations and the
renewal process distribution are unknown.  In this unknown sampling location
setup, it is shown that the mean-squared error in field estimation decreases as
$O(1/n)$ where $n$ is the average number of samples collected by the mobile
sensor. The average number of samples collected is determined by the
inter-sample spacing distribution in the renewal process.  An algorithm to
ascertain spatial field's bandwidth is detailed, which works with high
probability as the average number of samples $n$ increases. This algorithm works
in the same setup, i.e., in the presence of measurement-noise and unknown
sampling locations.
\end{abstract}


\IEEEpeerreviewmaketitle

\section{Introduction}
\label{sec:intro}

Consider a mobile sensor which has to acquire a spatially smooth field by moving
along a path or spatial trajectory~\cite{unnikrishnanVS2012,unnikrishnanVS2013}.
If the sensor is equipped with precise location information, high-precision
quantizers, and negligible measurement-noise, then the field reconstruction
process reduces to classical (noiseless) sampling and interpolation
problems~(see~\cite{papoulisE1966,jerrit1977,marvastin2001}). With
\textit{precise location information}, spatial field reconstruction or
estimation has been addressed in the presence of quantization as well as
measurement-noise~(for example,
see~\cite{pinskerO1980,masryT1981,dabeerKS2006,masryIF2009,wangID2009,kumarPE2013}).
In the context of spatial sampling with a mobile-sensor, a more challenging
setup is when the \textit{location of samples collected is not known}. Unknown
sampling locations is a \textit{fundamentally new topic} in spatial field
sampling and will be the central theme of this work.

The motivation for mobile-sampling without the knowledge of locations is
elucidated first. In practice, a device such as GPS (global positioning system)
or other elaborate distributed localization mechanisms can be used to localize a
sensor~\cite{patwariAKHMCL2005}. If a mobile-sensor's path and its velocity are
known, and if the mobile-sensor has an accurate clock, then the sampling
locations can be calculated from sample timestamps~\cite{unnikrishnanVS2012}.
However, all these elaborate mechanisms will add to the cost of mobile sensors,
and increase its recording overhead.  It would be desirable to get rid of the
timestamps, the GPS, the distributed mechanisms for localization and the
knowledge of velocity, and \textit{still} reconstruct the spatial field to a
desired accuracy. This is the core motivation behind the paper.

Spatially smooth, temporally fixed, and finite-support fields will be
considered in this work. The smoothness of spatial field will be
modeled by bandlimitedness.  It will be assumed that the mobile sensor
samples the field at locations obtained by an \textit{unknown} renewal
process. By unknown renewal process, we mean that the probability
distribution of the inter-sample locations and even the locations at
which field samples are obtained \textit{are not known}. It will also
be assumed that the field samples are affected by additive and
independent noise with zero mean and finite variance.  Except for
independence, zero-mean, and finite variance, it is assumed that the
\textit{noise distribution is also not known}. In such a challenging
setup, \textit{sampling rate} (corresponding to oversampling) will be
used in this work to decrease expected mean-squared error in
field-reconstruction. In other words, the mobile sensor will collect a
large number of readings on locations determined by an unknown renewal
process and estimates will be developed in this work to drive down the
expected mean-squared error in field-reconstruction with sampling
rate.

\begin{figure}[!htb]
\begin{center} 
\scalebox{1.0}{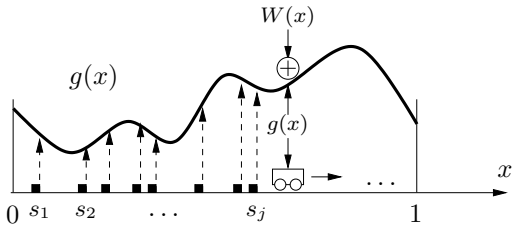}
\end{center}
\caption{\label{fig:intro} \sl \small The mobile sampling scenario under study
is illustrated. A mobile sensor, where the field is temporally fixed, collects
the spatial field's values at unknown locations $s_1$, $s_2$, $\ldots$, $s_m$.
It is assumed that $s_1, \ldots, s_m$ are realized from an arrival process with
an unknown renewal distribution to model the sensor's nonuniform velocity. It is
also assumed that the samples are affected by additive and independent
noise process $W(x)$. Our task is to estimate $g(x)$ from the readings $g(s_1) +
W(s_1), \ldots, g(s_m) + W(s_m)$.}
\end{figure}

The field sampling setup with a mobile sensor is illustrated in
Fig.~\ref{fig:intro}. To keep the analysis tractable, the spatial field is
assumed to be one-dimensional in space and temporally fixed in this first
exposition.\footnote{It is desirable to extend the analysis to multidimensional
fields which evolve with time. However, due to unknown sampling locations and
measurement-noise, the setup is already very challenging. For this reason, only
one-dimensional fields are considered in this work. It is fair to state that the
current work will be applicable to slowly changing spatial fields.} The mobile
sensor collects the spatial field's values at unknown locations $s_1$, $s_2$,
$\ldots$, $s_m$ derived from a \textit{renewal process with unknown inter-sample
distribution}~\cite{gallagerS2014}.

The spatial field measurements are affected by
additive and independent noise process $W(x)$, which has an unknown distribution
with zero-mean and finite variance. In this challenging setup, the goal is to
estimate the field from the readings collected by the mobile sensor.

Unknown location information on samples, (low-precision) quantization,
knowledge of the spatial field's bandwidth, and presence of additive
measurement-noise are four perils in the process of field estimation
(or reconstruction). In this work, leaving quantization issues aside,
the other three ailments will be addressed together in a
mobile-sampling framework. At first, a reconstruction method is
discussed where the bandwidth of the signal is known. Next, an
algorithm will be outlined in this work which handles an unknown (but
finite) bandwidth signal.  The main results shown in this work are as
follows:
\begin{enumerate}
\item For the field sampling setup illustrated in Fig.~\ref{fig:intro}
and with some regularity conditions on the inter-sample location
distribution, the expected mean-squared error in field estimation is
upper bounded as $O(1/n)$, where $n$ is the sampling density (or $n$
is the expected number of samples realized by the renewal process in
the interval of sampling). This result holds when zero-mean
finite-variance additive noise is present during sampling, and when
the sampling locations are obtained at unknown locations generated
from an unknown renewal process.
\item To address unknown bandwidth, an algorithm will be presented to
ascertain the correct bandwidth of the spatial field (with high
probability) in the mobile-sensor sampling paradigm. This algorithm
works in the presence of measurement-noise and when field samples are
obtained at unknown locations generated from an unknown renewal
process. The proposed algorithm requires the knowledge of mean and
variance of squared-noise.\footnote{If the measurement-noise is
stationary over space and time, as assumed, then its mean, variance
and other moments can be estimated conveniently by taking many
readings at the same location. This problem becomes more challenging
if the noise statistics vary with time. Also, the mean and variance of
squared noise refer to the second and fourth moment of a zero-mean
noise random variable.}
\end{enumerate}

\textit{Prior art:} The topic of unknown but uniformly distributed
sampling locations was recently introduced in the context of spatial
sampling in a finite interval~\cite{kumarO2015}. A detailed
differentiation with this earlier work is needed, since the setup and
results appear to be similar in nature. The estimates used are also
the same in the two works (see~\eqref{eq:agen}).  If $n$ sensors are
uniformly distributed in an interval, as in~\cite{kumarO2015}, then
their ordered (noise-affected) readings can be used to obtain a
bandlimited field estimate with mean-squared error of $O(1/n)$. The
derivation of this result utilizes established facts on the
mean-squared deviations of ordered uniformly distributed random
variables~\cite{kumarO2015,davidno2003}. It is also known that ordered
uniformly distributed random variables can be understood as
realizations of a Poisson renewal process~\cite{gallagerS2014}. In
contrast, in this current work the renewal process distribution is
assumed to be \textit{unknown}. This setup is much more challenging
than the previous work, since the distribution properties are assumed
to be unknown in the current work. As a result, the mean-squared error
analysis is derived from first principles. The final result on
mean-squared error analysis results in an upper bound of $O(1/n)$,
which is the same as in prior work in spite of unknown distribution
properties of the renewal process. One important aspect of result in
this work is that both the estimate and the mean-squared error result
are universal in nature (see~\eqref{eq:agen} and
Theorem~\ref{thm:ahatresult}). In summary, the sampling location's
distribution is assumed to be unknown, and is the major difference
with past work; and, the problem addressed in the current work is more
difficult!

Sampling and reconstruction of \textit{discrete-time} bandlimited signals from
samples taken at unknown locations was first studied by Marziliano and
Vetterli~\cite{marzilianoV2000}. This problem is addressed completely in a
discrete-time setup and solutions are combinatorial in nature.  A recovery
algorithm for bandlimited signals from a finite number of ordered nonuniform
samples at \textit{unknown sampling locations} has been proposed by
Browning~\cite{browningA2007}. This algorithm works in a deterministic setup.
Estimation of periodic bandlimited signals, where samples are obtained at
unknown locations obtained by a random perturbation of equi-spaced deterministic
grid, has been studied by Nordio et al.~\cite{nordioCVP2008}. More generally,
the topic of sampling with jitter on the sampling
locations~\cite{papoulisE1966},\cite[Chap.~3.8]{zayedA1993} is well known in the
literature.

This work is different from previous literature in the following non-trivial
aspects---(i) the sampling locations are generated according to an unknown
renewal process, where even the distribution defining the renewal process is
unknown; and, (ii) the field is affected by zero-mean additive independent noise
with finite variance, where the distribution of noise is also not
known.\footnote{It must be noted that renewal process generated sampling
locations are statistically dependent unlike in a sampling with jitter setup,
where the jitter values are typically assumed to be statistically independent.}


\noindent \textit{Notation:} Spatial fields which are temporally
fixed will be denoted by $g(x)$ and its variants.  The ${\cal
L}^\infty$-norm of a field $g(x)$ will be denoted by $\|g \|_\infty$.
The spatial derivative of $g(x)$ will be denoted by $g'(x)$.  The
number of (random) samples will be denoted by $M$, while $n$ will
denote the expected value of $M$.  Expectation will be denoted by
$\eE$. The set of integers, real numbers, and complex numbers will be
denoted by $\zZ$, $\rR$, and $\cC$, respectively. Finally, $j =
\sqrt{-1}$.

\noindent \textit{Organization:} Spatial field's model, the reconstruction
distortion criterion, the sampling model using a mobile sensor, and the noise
model on measurement (or field) is explained in Section~\ref{sec:fieldmodel}.
The estimation of spatial field with known bandwidth and samples taken on an
unknown renewal process is addressed in Section~\ref{sec:renewal}. An algorithm
for spatial field estimation with unknown bandwidth and samples taken on an
unknown renewal process addressed in Section~\ref{sec:algorithm}.
Measurement-noise is considered both in Section~\ref{sec:renewal} and
Section~\ref{sec:algorithm}. Simulation results are presented in
Section~\ref{sec:simulations}. Finally, conclusions will be presented in
Section~\ref{sec:conclusions}.

\section{Field model, distortion, sampling process by mobile sensor,
and measurement-noise model}
\label{sec:fieldmodel}

The models used for theoretical analysis presented in Section~\ref{sec:renewal}
and Section~\ref{sec:algorithm} will be discussed in this section. Field models
are discussed first.

\subsection{Field model}
It will be assumed that the field of interest is one dimensional, temporally
fixed, and spatially bandlimited.  Let $g(x)$ be the field, where $x \in \rR$ is
the spatial dimension. For reasons stated in Sec.~\ref{sec:intro}, it will be
assumed that the temporal variation of $g(x)$ is negligible, i.e., $g(x,t)
\equiv g(x)$ during the sampling process.
Temporally fixed assumption on the spatial field is \textit{suitable} when the
speed of the mobile sensor is much higher than the temporal rate of change of
the field (also see~\cite{unnikrishnanVS2012} for discussions). Temporally fixed
fields are more tractable to analysis, and this assumption will also help in
understanding the effect of location-unawareness on the field sampling process
in isolation.

Without loss of generality, it is assumed that $|g(x)| \leq 1$, $g(x)$ is
periodic and bandlimited with period equal to $1$, and has a bandwidth $2b\pi$,
where $b$ is a known positive integer. All these assumptions imply that $g(x)$
has a Fourier series in the interval $[0,1]$ with bounded Fourier series
coefficients
\begin{align}
g(x) & = \sum_{k = -b}^b a[k] \exp(j 2 \pi k x) \\
\mbox{ with  } a[k] & = \int_{0}^1 g(x) \exp(- j 2\pi k x)\mbox{d}x.
\label{eq:fseries}
\end{align}
Based on Bernstein inequality~\cite{hardylpi1959}, it follows that
\begin{align}
|g'(x)| \leq 2 b \pi \| g \|_\infty \leq 2 b \pi. \label{eq:bernstein}
\end{align}

\subsection{Distortion criterion}

For a first exposition, a mean-squared error will be used as the distortion
metric. If $\widehat{G}(x)$ is any estimate of the field $g(x)$, then the
distortion is defined as
\begin{align}
D & := \eE \left[ \int_{0}^1 \left| \widehat{G}(x) - g(x) \right|^2 \mbox{d}x
\right] \nonumber \\
& = \sum_{k = -b}^b \eE \left[ \left| A[k] - a[k] \right|^2 \right]
\label{eq:distortion}
\end{align}
where $A[k]$ is the Fourier series representation of $\widehat{G}(x)$. 

\subsection{Renewal process based sampling model}

It will be assumed that $X_1, X_2, \ldots$ be a renewal process which are the
separations between the sampling locations of a mobile sensor.\footnote{For a
renewal process, each $X_i > 0$.} The mobile sensor traveling in the interval
$[0,1]$ obtains samples at the locations $S_1:= X_1$, $S_2:=(X_1 + X_2)$,
$\ldots$, $S_M := (X_1 + X_2 + \ldots + X_M)$, where $M$ is the random number of
samples from the renewal process that fall in the interval $[0,1]$. The random
number of samples $M$ that will be obtained in $[0,1]$ satisfies a stopping rule
\begin{align}
X_1 + X_2 + \ldots X_M \leq 1 \mbox{ and } X_1 + X_2 + \ldots X_{M+1} > 1
\nonumber
\end{align}
and, therefore, is a well defined (measurable) random
variable~\cite{durrettp1996}. Let $X_1, X_2, X_3, \ldots$ have the common
distribution $X$.  For analysis purposes, it will be assumed that 
\begin{align}
0 < X \leq \frac{\lambda}{n} \mbox { and } \eE (X)  = \frac{1}{n}
\end{align}
where $\lambda > 1$ is a finite constant independent of $n$. In other words, $0
< nX \leq \lambda$, or $n X$ is a bounded random variable.  The random variable
$M$ is the (random but large) number of readings made to determine the field
$g(x)$ by a mobile sensor, which is unaware of its sampling locations. The main
help in field reconstruction is from the parameter $n$. It will be shown that as
the parameter $n$ becomes large, there exists an estimate of the field which
converges to the true field $g(x)$  with respect to the distortion
in~\eqref{eq:distortion}.  Our estimate will work \textit{without the knowledge}
of (exact values of) $X_1, X_2, \ldots, X_M$, and more ambitiously the
distribution of $X$ will \textit{also be unknown}. It is assumed however that
the distribution of $nX$ has a non-zero support only on $(0,\lambda]$.

\subsection{Measurement-noise model}

It will be assumed that the field samples are affected by additive
measurement-noise; that is, $g(x) + W(x)$ is sampled, where $W(x)$ is
additive and independent.  The measurement-noise process is
independent, which means for any positive integer $l$, $W(x_1),
W(x_2), \ldots, W(x_l)$ are i.i.d.~for distinct $x_1, x_2, \ldots,
x_l$. Observe that $W(x)$ is not a continuous-time white noise
process, but its sampled version will be a discrete-time white noise
process.  It is also assumed that $W(x)$ is independent of the
renewal-process $X_1, X_2, X_3, \ldots$ which generates the sampling
locations.

\section{Field estimation from samples obtained on an unknown renewal process}
\label{sec:renewal}

In this section, the Fourier series of the periodic bandlimited field $g(x)$
will be estimated by observing measurement-noise affected samples obtained on an
unknown renewal process. As outlined in Section~\ref{sec:intro} and
Section~\ref{sec:fieldmodel}, the term unknown renewal process refers to unknown
sampling locations and unknown inter-sample spacing distribution.

Recall that the sampling locations are defined as
\begin{align}
S_1 & = X_1, \nonumber \\
S_2 & = X_1 + X_2 = S_1 + X_2, \nonumber \\
\mbox{and } S_M & = X_1 + \ldots + X_M =  S_{M-1} + X_M.
\end{align}
The sampling locations $S_1, S_2, \ldots, S_M$ are unknown, as well as with
unknown distribution. Noise-affected field values $g(S_1)+W(S_1), g(S_2)+W(S_2),
\ldots, g(S_M)+W(S_M)$ are available for the estimation of the field $g(x)$. The
key to our estimation procedure will be the field's Fourier series coefficient
estimate $\widehat{A}_{\gen}[k]$ defined by
\begin{align}
\widehat{A}_{\gen}[k] & := \frac{1}{M} \sum_{i = 1}^M \left\{g(S_i) +
W(S_i)\right\} \exp \left(- \frac{j 2\pi k i}{M} \right) \label{eq:agen}
\end{align}
where $-b \leq k \leq b$. This estimate works without the knowledge of $S_1,
\ldots, S_M$ since the noise-affected field values $g(S_1) + W(S_1), \ldots,
g(S_M) + W(S_M)$ are the samples recorded by the mobile sensor in our model.
This formula is a Riemann-sum like approximation to the Fourier series integral
formula in \eqref{eq:fseries} with two assumptions: (i) the sample locations
given by $S_i, 1 \leq i \leq M$ are ``near'' the grid points $i/M, 1 \leq i \leq
M$; and (ii) the measurement-noise part in \eqref{eq:agen} averages out to
``near-zero''.  The effect of these assumptions is analyzed; and, in the
estimation of Fourier series coefficients from $\widehat{A}_{\gen}[k]$ (in the
mean-squared sense) the following theorem is noted.
\begin{theorem}
\label{thm:ahatresult}
Let $\widehat{A}_{\gen}[k]$ be as defined in \eqref{eq:agen}. Let $X_i$ be
i.i.d.~positive (inter-sample spacing) random variables such that $\eE(X_i) =
1/n$, and distribution of $X_i$ has support in $(0, \lambda/n]$.  Let $W(x)$ be
a measurement-noise process independent of $X_i, i \in \zZ$ with zero-mean and
finite variance. Then
\begin{align}
\eE \left[ \left| \widehat{A}_{\gen}[k] - a[k] \right|^2 \right] \leq
\frac{C}{n}
\end{align}
where $C$ is a constant that is independent of $n$, and depends on renewal
process parameter $a$ and the signal bandwidth parameter $b$.  Correspondingly
the distortion in \eqref{eq:distortion} is bounded as $D_{\gen} \leq (2b + 1)
C/n$ or $D_{\gen} = O(1/n)$. 
\end{theorem}

\proof To maintain the flow of the results, the key ideas and inequalities in
the proof will be proved in this section while the technically detailed
(mundane) statistical calculations will be presented in
Appendix~\ref{ap:samplinggrid} and Appendix~\ref{ap:riemannapprox}.

The estimate in \eqref{eq:agen} has been designed with the assumption that
$g(S_1)$ has been observed at $t= 1/m$, $g(S_2)$ has been observed at $t = 2/m$,
and $g(S_i)$ has been observed at $t = i/m$ for various values of $i$. This is
our \textit{key statistical approximation}. The mean-squared error in making
this approximation will be analyzed next. The estimate $\widehat{A}_{\gen}[k]$
consists of two conceptual parts
\begin{align}
\widehat{A}_{\gen}[k] & = \underbrace{\frac{1}{M} \sum_{i = 1}^M g(S_i) \exp
\left(- \frac{ j 2\pi k i}{M} \right)}_{\widehat{A}[k]} + \nonumber \\
& \mbox{\hspace{2cm}} \underbrace{\frac{1}{M} \sum_{i = 1}^M W(S_i) \exp \left(-
\frac{j 2\pi k i}{M} \right)}_{\wAvg[k]} \label{eq:ahatgen}
\end{align}
where the first and second terms correspond to the signal and
the measurement-noise part,
respectively. These terms will be analyzed separately. Let 
\begin{align}
A_{R}[k] := \frac{1}{M} \sum_{i = 1}^M g(i/M) \exp(- j 2 \pi k i/M)
\end{align}
be the $M$-point Riemann approximation of $a[k]$ in \eqref{eq:fseries}.  The
approximation $A_{R}[k]$ is random due to presence of $M$. 
Since $|a_1 + a_2|^2 \leq 2|a_1|^2 + 2|a_2|^2$, therefore
\begin{align}
\eE \left[ \left| \widehat{A}_{\gen}[k] - a[k] \right|^2 \right] = 2 \eE
\Bigg[\Big|\widehat{A}[k] & - a[k] \Big|^2 \Bigg]  + \nonumber \\
& \mbox{\hspace{0.3cm}} 2 \eE \left[ |\wAvg[k]|^2 \right].
\label{eq:msebreakdown}
\end{align}
Next, from the triangle inequality~\cite{rudinp1976}, 
\begin{align}
\left|\widehat{A}[k] - a[k] \right| \leq |\widehat{A}[k] - A_{R}[k]| + |A_{R}[k]
- a[k]|
\end{align}
and $|a_1 + a_2|^2 \leq 2|a_1|^2 + 2|a_2|^2$, it follows that
\begin{align}
\eE\left[ \left|\widehat{A}[k] - a[k] \right|^2 \right] \leq 2 \eE \Big[
|\widehat{A}[k] - & A_{R}[k]|^2 \Big] + \nonumber \\
& 2 \eE\left[ |A_{R}[k] - a[k]|^2 \right] \label{eq:ahatbound}
\end{align}
The term $|A_{R}[k] - a[k]|$ will be bounded using the smoothness properties of
$g(x)$ and the mean-value theorem; and, the mean-squared value of
$|\widehat{A}[k] - A_{R}[k]|$ will be upper-bounded using eaxchangeability of
$X_1, X_2, \ldots, X_M$ conditioned on the stopping time
$(M+1)$~\cite{durrettp1996}.  The next parts are devoted to these analyses.
First, $|\widehat{A}[k] - A_{R}[k]|$  is considered.  Note that
\begin{align}
|\widehat{A}[k] & - A_{R}[k]| \nonumber \\
& = \Bigg| \frac{1}{M} \sum_{i = 1}^M g(S_i) \exp \left(- j 2\pi k \frac{i}{M}
\right) - \nonumber \\
& \mbox{\hspace{2cm}} \frac{1}{M} \sum_{i = 1}^M g \left(\frac{i}{M}\right) \exp
\left(- j 2\pi k \frac{i}{M} \right) \Bigg| \\
& \leq  \frac{1}{M} \sum_{i = 1}^M \left| g(S_i) - g\left(\frac{i}{M}\right)
\right| \label{eq:ahatar}
\end{align}
where the last step using the triangle inequality~\cite{rudinp1976}.  Since
$(a_1 + \ldots + a_m)^2 \leq m (a_1^2 + \ldots + a_m^2)$, for any real numbers
$a_1, \ldots, a_m \in \rR$ and any integer $m$, so
\begin{align}
|\widehat{A}[k] & - A_{R}[k]|^2 \nonumber \\
& \leq  \frac{1}{M^2} M \sum_{i = 1}^M \left| g(S_i) - g\left(\frac{i}{M}\right)
\right|^2 \\
%
%
& \leq  \| g' \|^2_\infty \frac{1}{M} \sum_{i = 1}^M \left| S_i - \frac{i}{M}
\right|^2, \label{eq:gtoepsilon}
\end{align}
where the last step uses $|g(x_1) - g(x_2)| \leq \| g'\|_\infty |x_1 - x_2|$ for
any $x_1, x_2 \in [0,1]$. Taking expectations in \eqref{eq:gtoepsilon},
\begin{align}
\eE\left[ |\widehat{A}[k] \right. & \left. - A_{R}[k]|^2 \right] \nonumber \\
& \leq  \| g' \|^2_\infty \eE\left[ \frac{1}{M} \sum_{i = 1}^M \left| S_i -
\frac{i}{M} \right|^2 \right]. \label{eq:ahatarmse}
\end{align}
From Appendix~\ref{ap:samplinggrid}, it is noted that
\begin{align}
\eE\left[ \frac{1}{M} \sum_{i = 1}^M \left| S_i - \frac{i}{M}
\right|^2 \right] \leq \frac{C_1}{n} \label{eq:epsilonFn}
\end{align}
for any $F$ and as $n$ becomes large, and the constant $C_1 > 0$ depends
on $\lambda$ and is independent of $n$.

For $a[k] - A_{R}[k]$, it is shown in Appendix~\ref{ap:riemannapprox} that
\begin{align}
|a[k] - A_{R}[k]| & \leq \frac{C_2}{M} \\
\mbox{or } \eE \left[ |a[k] - A_{R}[k]|^2 \right] & \leq \eE\left(
\frac{C_2^2}{M^2} \right) \leq \frac{C_2 \lambda^2}{n^2}. \label{eq:akarmse}
\end{align}
The constant $C_2 > 0$ depends on the field's bandwidth parameter $b$ and is
independent of $n$.

Finally, the mean-squared value of $\wAvg[k]$ has to be characterized.
For this part, note that
\begin{align}
\eE \left( |\wAvg[k]|^2\right) & = \eE \left| \left\{ \frac{1}{M} \sum_{i = 1}^M
W(S_i) \exp \left(- \frac{j 2\pi k i}{M} \right) \right\} \right|^2 \nonumber \\
& = \eE \Bigg[ \frac{1}{M^2} \sum_{i, l = 1}^M W(S_i) W(S_l) \times \nonumber \\
& \mbox{\hspace{1cm}} \exp \left(-
\frac{j 2\pi k i}{M} \right)  \exp \left( \frac{j 2\pi k l}{M} \right) \Bigg] \\
& \stackrel{(a)}{=} \eE \left[ \frac{1}{M^2} \sum_{i = 1}^M |W(S_i)|^2 \right]
\\
& = \eE\left[ \frac{\sigma^2}{M} \right] \leq \frac{\sigma^2 \lambda}{n} .
\label{eq:noisemse}
\end{align}
The equality in $(a)$ follows since the noise process $W(x)$ is independent,
$S_1, S_2, \ldots, S_M$ are distinct, and $M$ (which depends on sampling
process) is independent of the measurement-noise process $W(x)$.

Putting together results from \eqref{eq:ahatarmse}, \eqref{eq:epsilonFn},
\eqref{eq:akarmse}, and \eqref{eq:noisemse} in \eqref{eq:msebreakdown},
\begin{align}
\eE \Big[ & |\widehat{A}_{\gen}[k] - a[k]|^2 \Big] \\
& \leq 4 \| g' \|_\infty^2 \frac{C_1}{n} + 4 \frac{C_2^2 \lambda^2}{n^2} +
2 \frac{\sigma^2 \lambda}{n} \\
& \leq \left( 4 (2b+1)^2 \pi^2 C_1 + \frac{4 C_2^2 \lambda^2}{n} + 2 \sigma^2
\lambda \right) \frac{1}{n} \\
& \leq \frac{C}{n}
\end{align}
for some $C > 0$ which does not depend on $n$. Observe that the constant $C$
becomes larger with larger bandwidth ($b$), larger noise variance ($\sigma^2$),
and a larger spread of renewal distribution ($\lambda$). A larger value of
$\lambda$, indicates that the (unknown) sampling locations are more spread-out
around their mean value of $1/n$. This results in a worse proportionality
constant. The main result of the theorem is complete. \endproof

The following remarks are useful in the context of above result.

\begin{remark}
\label{rem:tightness}
The estimate works on the principle that $S_i$ is nearly
equal to $(i/M)$, and statistical averaging (with large $n$) is expected to help
in convergence of $\widehat{A}[k]$ towards $a[k]$. The exact rate of convergence
is shown to be $O(1/n)$. The simulations presented in
Section~\ref{sec:algorithm} illustrate a mean-squared error proportional to
$1/n$. This suggests that the $O(1/n)$ upper-bound is tight, at least for the
estimate presented in \eqref{eq:ahatgen}.
\end{remark}

\begin{remark}
\label{rem:poissondist}
It was assumed that the distribution of $nX$ has a support in
$(0, \lambda]$, where $\lambda$ is a finite constant. This assumption simplifies the
analysis but omits distribution with infinite support such as the exponential
distribution. In the special case when $n X$ is exponentially distributed,
conditioned on $M$, the random variables $S_1, S_2, \ldots, S_M$ will correspond
to ordered $\mbox{Unif}[0,1]$ random variables. In case if $nX$ is exponentially
distributed, the mean-squared error between $\widehat{A}_{\gen}[k]$ and $a[k]$
can be shown to be $O(1/n)$ by using existing results in the
literature~\cite{kumarO2015}. 
\end{remark}

\begin{remark}
\label{rem:arrivaldist}
From the proof in Appendix~\ref{ap:samplinggrid}, the $O(1/n)$ decay in
distortion will hold if $\eE(1/M)$ decreases proportionally to $1/n$ and
$\eE(M+1)$ is proportional to $n$. The latter condition can be established
easily by Wald's identity~\cite{durrettp1996}. The former condition will need
some sophisticated statistical analysis with stopping-times and has been left as
a future work.  At a high level, $\eE(1/M)$ can be expected to decrease as
$O(1/n)$ since $(M/n) = 1$ almost-surely as the sampling rate $n$ increases.
The assumption that $nX \leq \lambda$ makes the mean-squared error analysis a
little convenient.
\end{remark}

\begin{remark}
\label{rem:pontogram}
Renewal process with small mean (of $1/n$) result in a
`pontogram', which is connected to the Brownian Bridge~\cite{steinebachZO1993}.
In the spatial-sampling context, if $M(x)$ is  the number of samples taken up
till location $x$ (with $M(1) = M$ in this work's notation), then $\sqrt{n}[M(x)
- x M(1)]$ will be a generalized pontogram as a function of $x \in [0,1]$. Then,
it is known that the worst deviation of the Pontogram from a Brownian bridge is
negligible (with high probability) as $n$
increases~\cite[Theorem~2.1]{steinebachZO1993}. This indicates that, in the
limit of $n$ large, the mobile sensor will be sampling the spatial field on a
Brownian bridge! The properties of a bandlimited field being observed on a
Brownian bridge (at unknown points) is an interesting topic of study for future
research.

Like in a Brownian bridge, this result also suggests that the variance of
sample-locations in the middle is larger than those at the edges of the sampling
interval. In the future, it would be interesting to design estimates of $a[k]$
which utilize this property.
\end{remark}

\begin{remark}
\label{rem:orthobasis}
An inspection of proof of Theorem~\ref{thm:ahatresult} suggests that its result
will hold if $g(x)$ is a field in any set of bounded dynamic-range fields having
an orthonormal basis, and having smoothness properties such as finite derivative
over the entire class. Bandlimitedness would translate to having finite degrees
of freedom (or finite number of non-zero coordinates in the orthonormal basis).
Orthonormal basis would imply that the degrees of freedom can be obtained using
a suitable inner product, which can be approximated using a Riemann sum
(see~\eqref{eq:fseries} and~\eqref{eq:ahatgen}). Smoothness properties of the
set of fields will enable counterparts of \eqref{eq:bernstein} required for
approximation analysis (see~\eqref{eq:gtoepsilon}). This generalization, we
believe, is analogous to the Fourier series development followed in this work.
This generalization is not established in the current work due to space
constraints, and more importantly for simplicity of exposition.
\end{remark}

\section{Bandwidth determination using field samples obtained on an unknown
renewal process}
\label{sec:algorithm}

In some applications, the bandwidth or the essential bandwidth of the spatial
field may not be known~\cite{slepianO1976}. This can be because the essential
bandwidth of spatial fields change with time~\cite{ranieriVS2013}, or because
the field being observed is not characterized for bandwidth, or because the
sampling path is not a straight line.\footnote{The one-dimensional ``slices'' of
a two-dimensional field have a bandwidth that depends on the
slicing~\cite{marksI1990}.} Under some technical assumptions, an algorithm is
outlined to find the bandwidth $b$ of the field in an asymptotic setting where
the sampling rate $n$ increases asymptotically.

Consider a spatial field $g(x)$ which has a finite but unknown bandwidth
parameter $b$. From \eqref{eq:ahatgen} and the result in
Theorem~\ref{thm:ahatresult}, $\widehat{A}_{\gen}[k]$ for $|k| > b$ will
converge (in the mean-squared sense, and therefore in probability) to zero.
This observation can be used to design a reconstruction algorithm for a
spatially bandlimited process with unknown bandwidth. The following assumptions
are made for this section:
\begin{enumerate}
\item The spatial field has a finite but unknown bandwidth parameter $b$ (see
\eqref{eq:fseries}).
\item All the non-zero Fourier series coefficients are \textit{larger than}
$\Delta$ in magnitude, where $\Delta >0$ is a constant.
\item The measurement-noise is zero-mean, its second moment (variance) is known,
and its fourth moment is finite (i.e., $\eE(W^4) < \infty$). 
\end{enumerate}
A non-zero constant $\Delta$ (in the second assumption above) is needed to
ascertain the bandwidth, since any statistical estimate  (such as
$\widehat{A}_{\gen}[k]$) will be negligible outside the bandwidth $b$ but not
exactly zero. A threshold parameter $\Delta$ ensures that the sampling rate $n$
can be made large enough to get rid of negligible but otherwise non-zero Fourier
series coefficients.\footnote{As expected, for smaller $\Delta$, a larger
sampling rate $n$ will be required.}

For any $k$, an estimate for $a[k]$ can be obtained from $\widehat{A}_{\gen}[k]$
in \eqref{eq:ahatgen}. The tricky part is determination of $b$, that is, when to
stop the Fourier series coefficient estimation! In other words, a stopping
condition is needed.  The next paragraph summarizes this stopping condition and
then an algorithm for bandwidth determination is presented, with a sketch of its
technical correctness.

Consider a simplified problem, where a bandlimited but unknown bandwidth signal
$h(x)$ is available. It is known that the bandwidth of $h(x)$ is finite, but its
value is not known. To reconstruct the field, the bandwidth of $h(x)$ is
required.  Let $c[k]$ be the Fourier series of $h(x)$. The Fourier series
coefficients $c[0], c[1], c[-1], \ldots$ can be sequentially computed.  The main
issue is when the Fourier series coefficients computation should be stopped?  To
this end, note that since $h(x)$ is available so is its energy $\int_{0}^1
|h(x)|^2 \mbox{d}x$. By Parseval's theorem, it is known that 
\begin{align}
\int_{0}^1 |h(x)|^2 \mbox{d}x = \sum_{k = -\infty}^{\infty} |c[k]|^2.
\label{eq:parsevals}
\end{align}
So, $c[0], c[1], c[-1], \ldots$ can be computed till \textit{the energy} in the
collected coefficients \textit{matches} with that of $h(x)$. Since the energy of
$h(x)$ is finite, and $b$ is finite by assumption, so this process will end in
$(2b+1)$ number of steps. An adaptation of this idea will be used in the
stochastic sampling setup with a mobile sensor.  Since \textit{accurate
approximations} of the field's energy and Fourier series coefficients are
available only for $n \rightarrow \infty$, so an approximate adaptation of this
algorithm is needed to address finite but large values of $n$.

An estimate of spatial field energy (see~\eqref{eq:parsevals}) is needed since
the field is not available in entirety but only through noise-affected samples
at unknown locations. The spatial field's energy estimate is \textit{defined} as 
\begin{align}
E_g := \frac{1}{M} \sum_{i = 1}^M [g(S_i) + W(S_i)]^2 - \sigma^2,
\label{eq:empenergy}
\end{align}
where $\sigma^2$ is the noise variance and is assumed to be known. The intuition
in the above estimate is that $W(x)$ and $X_1^M, M$ are independent (and hence
uncorrelated), and $W^2(S_i), i = 1, \ldots, M$ will average near $\sigma^2/M$.
The analysis in Appendix~\ref{ap:energyestimate} shows that the mean-squared
value of $\left|E_g - \int_{0}^1 g^2(x) \mbox{d}x \right|$ is bounded as
\begin{align}
\eE\left[ \Big|E_g - \int_{0}^1 g^2(x) \mbox{d}x \Big|^2 \right] & \leq
\frac{\tilde{C}}{n} \label{eq:egconvmain}
\end{align}
where $\tilde{C} > 0$ is some constant independent of $n$ and depends on $g(x)$
only through $b$. Therefore $E_g$ converges to the field energy $\int_{0}^1 g^2
(x) \mbox{d}x$ in the mean-squared sense (and hence in probability).

As sampling rate $n$ becomes larger, the empirical energy in
\eqref{eq:empenergy} converges in mean-squared sense to the true energy of
$g(x)$.  An estimate for Fourier series coefficients has been presented in
\eqref{eq:ahatgen}, which converges in mean-squared sense as sampling rate 
$n$ becomes large. If this estimate is below the threshold $\Delta$ by some
margin, the Fourier series coefficient can be set to zero by a thresholding
operation. For asymptotic $n$, this process will result in (mean-squared)
correct Fourier series coefficients. Similarly, as noted earlier, each
$\widehat{A}_{\gen}[k]$ converges in mean-squared sense to the correct $a[k]$.
This motivates the following estimation algorithm, if each Fourier series
coefficient is more than $\Delta$ in magnitude, where $\Delta > 0$ is a positive
parameter.
\begin{enumerate}
\item Calculate an estimate $E_g$ for the spatial field's energy as in
\eqref{eq:empenergy}.
\item Start with $B = 0$. 
\item Calculate the Fourier series coefficient $\widehat{A}_{\gen}[B]$ and
$\widehat{A}_{\gen}[-B]$ as in \eqref{eq:ahatgen}. If the estimates are more
than $\Delta - \frac{1}{\cbrt{n}}$ in magnitude, retain them. Otherwise, set
$\widehat{A}_{\gen}[B]$ and $\widehat{A}_{\gen}[-B]$ as zero. It will be shown
shortly that non-zero coefficients succeed while zero coefficients fail in this
test with high probability, as $n$ increases asymptotically.
\item Increase $B$ by $+1$. Repeat the process in previous step till 
\begin{align}
- \frac{\Delta^2}{2} \leq \sum_{k = -B}^B \left| \widehat{A}_{\gen}[k] \right|^2
- E_g \leq \frac{\Delta^2}{2}. \label{eq:approxstop}
\end{align}
It will be shown shortly that this test will be met \textit{only} by the correct
bandwidth $b$ with high probability, as $n$ increases asymptotically.
\end{enumerate}
Claim in Item~3) above follows by Chebychev inequality~\cite{durrettp1996}. If
$|a[k]| \geq \Delta$, it is noted that
\begin{align}
\pP\left[ |\widehat{A}_{\gen}[k] - a[k]| < 1/\cbrt{n} \right] & \leq \eE
\frac{\left| \widehat{A}_{\gen}[k] - a[k] \right|^2 }{ (1/\cbrt{n})^2 } \\
& \leq \frac{C}{n^{1/3}} \label{eq:agenakinprob}
\end{align}
which means that $|\widehat{A}_{\gen}[k] - a[k]| < 1/\cbrt{n}$ with high
probability. Since $|a[k]| \geq \Delta$ by assumption, so 
\begin{align}
\Delta - \frac{1}{\cbrt{n}} \leq |a[k]| - |a[k] - \widehat{A}_{\gen}[k]| \leq
|\widehat{A}_{\gen}[k]|
\end{align}
with high probability. This establishes (the obvious) that
$\widehat{A}_{\gen}[k]$ will have a magnitude greater than $\Delta - 1/\cbrt{n}$
with high probability if $a[k]$ has a magnitude greater than $\Delta$. 

By similar argument, if $a[k] = 0$, then 
\begin{align}
\pP\left[ |\widehat{A}_{\gen}[k] - a[k]| < \Delta - \frac{1}{\cbrt{n}} \right] &
\leq \eE \frac{\left| \widehat{A}_{\gen}[k] - a[k] \right|^2 }{(\Delta -
n^{1/3})^2} \nonumber \\
& \leq \frac{C}{n (\Delta - n^{1/3})^2} \label{eq:agenzeroinprob}
\end{align}
which converges to zero with increasing $n$. So, if $a[k] = 0$, then
$\widehat{A}_{\gen}[k]$ is smaller than $\Delta - 1/\cbrt{n}$ with high
probability. 

Claim in Item~4) follows by Chebychev inequality and finiteness of $B$. From
\eqref{eq:egconvmain} and Chebychev inequality, it follows that
\begin{align}
\pP \left[ \left| E_g - \int_{0}^1 g^2(x) \mbox{d}x \right| > \frac{1}{\cbrt{n}}
\right] \leq \frac{\tilde{C}}{\cbrt{n}} \label{eq:egconvhp}
\end{align}
or with high probability or $E_g$ is close to $\int_{0}^1 g^2(x)\mbox{d}x$.  The
following inequalities are noted, each of which holds with high probability.
From \eqref{eq:agenakinprob} and \eqref{eq:agenzeroinprob},
\begin{align}
a[k] \neq 0, & ||\widehat{A}_{\gen}[k]|^2 - |a[k]|^2| \leq O(1/\cbrt{n}) \\
a[k] = 0, & |\widehat{A}_{\gen}[k]| = 0
\end{align}
where the second equality is achieved by thresholding the near-zero
$|\widehat{A}_{\gen}[k]|$ against $\Delta - 1/\cbrt{n}$.  That is, each
estimated $\widehat{A}_{\gen}[k]$ (for $-B \leq k \leq B$) is equal to zero
$a[k]$ or at a maximum distance of $1/\cbrt{n}$ to a non-zero $a[k]$.  So the
maximum difference between coefficient energies is
\begin{align}
\sum_{k = -B}^B |\widehat{A}_{\gen}[k]|^2  - \sum_{k = -B}^B |a[k]|^2 = O\left(
\frac{1}{\cbrt{n}} \right). \label{eq:agenakenergygap}
\end{align}
By Parseval's relation,
\begin{align}
\sum_{k = -B}^B |a[k]|^2 = \int_{0}^1 g^2(x)\mbox{d}x \label{eq:parsevalsagain}
\end{align}
where the maximum number of nonzero $a[k]$ is $\left[ \int_{0}^1 g^2(x)
\mbox{d}x\right]/\Delta^2$, since  $|a[k]| \geq \Delta$ by assumption. Finally,
the maximum difference between energy estimate of the field and the actual
energy is 
\begin{align}
\left| E_g - \int_{0}^1 g^2(x) \mbox{d}x \right| \leq \frac{1}{\cbrt{n}}
\label{eq:energygap}
\end{align}
From \eqref{eq:agenakenergygap}, \eqref{eq:parsevalsagain}, and
\eqref{eq:energygap}, it follows that 
\begin{align}
E_g - \sum_{k = -B}^B |\widehat{A}_{\gen}[k]|^2 = O(1/\cbrt{n})
\end{align}
only for the correct value of bandwidth $B$. Therefore, the stopping condition
in \eqref{eq:approxstop} will be met with high probability.

\begin{remark}
\label{rem:thresholds}
It must be noted that the above algorithm works for
asymptotic $n$ and convergence rate guarantees are not given in this first
exposition. It is possible to select some other function of $n$ instead of
$1/\cbrt{n}$ in Item~3) and some other threshold than $\Delta^2/2$, and optimize
the probability of error for a finite (but large) $n$. This analysis is left for
future work, but simulations based on the above algorithm will be presented.
\end{remark}

\section{Simulation results}
\label{sec:simulations}

\begin{figure*}[!htb]
\centering
\includegraphics[scale=0.20]{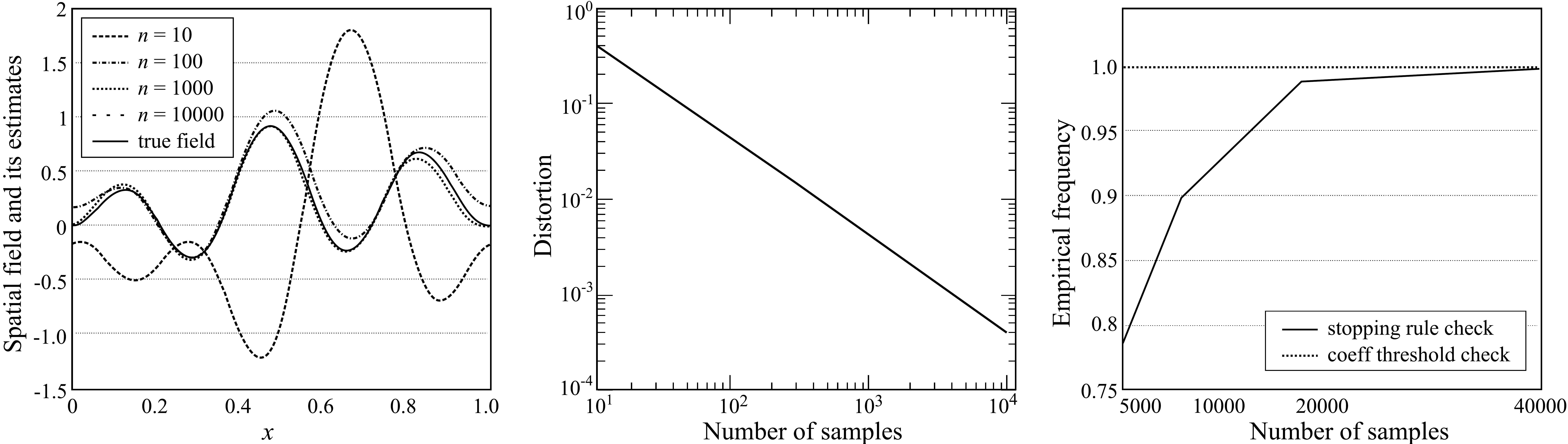}
\caption{\label{fig:MSEPlots} For the first two plots, the Fourier series
coefficients of the field are given in~\eqref{eq:plot12Fseries}.  In the first
plot, the convergence of random realizations of $\widehat{G}(x)$ to $g(x)$ is
illustrated. The convergence with increasing values of $n$ is apparent. The
field estimate for $n = 10000$ is nearly identical to the true field (and is not
visible in the plot). In the second plot the distortion, averaged over $10000$
random trials, is shown. The distortion decreases as $O(1/n)$ as expected
(see~Theorem~\ref{thm:ahatresult}).  In the third plot, it is observed that the
stopping rule check and coefficient threshold check are met successfully if $n$
exceeds $20000$. The plotted numerical values will depend on the value of
$\Delta$.}
\end{figure*}

Some simulation results are presented in Fig.~\ref{fig:MSEPlots}. In
these simulations, additive measurement-noise was generated using
Uniform$[-1,1]$ random variables. The random sampling locations were
obtained using a renewal process with uniform inter-sample spacing
distribution. The knowledge of this distribution, as explained in
\eqref{eq:ahatgen} and Section~\ref{sec:renewal}, was not used in the
field reconstruction. The field was generated in different ways for
the first and second plots, and the third plot. This is explained
next.

In the first and second plots, which evaluates the performance of
field estimate in Section~\ref{sec:renewal} for field with known
bandwidth, it is assumed that there is a field $g(x)$ with Fourier
series coefficients given by
\begin{align}
a[0] = {0.2445}, a[1] = - 0.0357 + j 0.0478 \nonumber \\
a[2] = 0.0978+j0.0729, a[3] = - 0.1796 - j0.0756 \nonumber \\
a[k] = \bar{a}[-k], \mbox{ and } a[k] = 0 \mbox{ for } |k| > 3.
\label{eq:plot12Fseries}
\end{align}
These Fourier series coefficients were obtained by independent trials
of Uniform$[-1,1]$ random variables (to obtain the real and imaginary
parts). Conjugate symmetry $a[k] = \bar{a}[-k]$ ensures that the field
is real-valued. Finally, the field was scaled to limit its dynamic
range within $[-1,1]$.

For the third plot, which evaluates the bandwidth estimation algorithm
of Section~\ref{sec:algorithm}, it is assumed that there is a field
$g(x)$ with Fourier series coefficients given by
\begin{align}
a[0] &= 0.1, a[1] = -0.1, a[12] = 0.1, \nonumber \\
a[k] &= \bar{a}[-k], \mbox{ and } a[k] = 0 \mbox{ otherwise}. \label{eq:plot3Fseries}
\end{align}
In these simulations $\Delta$ is set as $0.1$ Larger values of
$\Delta$ are more desirable for the algorithm proposed in this
section. Coefficient threshold check tests whether the non-zero
coefficients in the Fourier series of $g(x)$ are estimated as more
than $\Delta - 1/\cbrt{n}$ and the zero coefficients are $0$ (see
Item~3) in the algorithm above). The stopping rule in
\eqref{eq:approxstop} requires that the estimated field energy is
within $\Delta^2/2 = 0.05$ of the original.  This requirement is very
stringent and as a result, the stopping rule condition is violated
(never or wrongly met) for smaller values of $n$. For $\Delta = 0.1$,
the threshold $\Delta - 1/\cbrt{n}$ is positive only when $n \geq
1000$. For this reason, simulation begins from $n = 5000$ in the third
plot. 

In the first plot, the convergence of random realizations of
$\widehat{G}(x)$ to $g(x)$ can be observed with increasing $n$. The
graph for $n = 10000$ is near identical to the true field and cannot
be seen in the graph. The mean-squared error, averaged over $10000$
random trials, decreases as $O(1/n)$ as illustrated in the second
(log-log) plot. This is in consonance with our results in
Theorem~\ref{thm:ahatresult}. Finally, in the third plot, it is
observed that the stopping rule check and coefficient threshold check are
met successfully if $n$ exceeds $20000$. These numerical values will
change depending on the value of $\Delta$.

\section{Concluding remarks}
\label{sec:conclusions}

This work introduced the estimation of bandlimited spatial fields from
noise-affected samples taken at \textit{unknown sampling locations}, where the
locations are generated from a renewal process with unknown distribution.
Sampling rate was used to combat against additive measurement-noise as well as
unknown sampling locations. A spatial field estimate, which converges to the
true spatial field in the mean-squared sense at the rate
$O(1/\mbox{sampling rate})$, was presented and its analysis was the first main
result of the work. A spatial field bandwidth determination algorithm from field
samples collected at unknown sampling locations, which works with probability
one as sampling rate increases asymptotically, was proposed and its correctness
was the second main result of this work. Simulation results, consonant with the
theoretical analysis, were also presented. 

This work opens a flurry of interesting ideas related to reconstruction of
spatial fields from samples collected at unknown sampling locations.  How can
the field estimates be derived (developed) when the renewal process distribution
or the noise distribution is known? Is the distortion result developed in this
paper optimal in certain circumstances? How will the distortion change in the
presence of sample quantization? How should the field estimation change if a
fraction of field samples are taken at known locations or without
measurement-noise or both? What field estimation or reconstruction strategy
should be used to tackle the sampling of non-bandlimited fields? In all these,
and many more, questions we expect sampling rate to play a fundamental role in
the obtained answers.

\appendices

\section{Mean-squared closeness of renewal-process sampling grid to a uniform
grid}
\label{ap:samplinggrid}

For analysis purposes, let
\begin{align}
R_M = 1 - (X_1 + X_2 + \ldots X_M)
\end{align}
be the remaining distance between the last location of sampling and the end of
field support (or sampling vehicle's terminal stop). Observe that $R_M$ is
bounded since 
\begin{align}
R_M \leq X_{M+1} \leq \frac{\lambda}{n}.
\end{align}
First the average value of $X_M$ will be determined, conditioned on $M = m$.
Since $S_M + R_M = 1$ by definition, so
\begin{align}
\eE(S_M & + R_M | M = m)  = 1 \\
\mbox{or } m \eE(X_1|M = m) & + \eE(R_M|M = m) = 1 \label{eq:exch} \\
\mbox{i.e., } \eE(X_1|M = m)  & = \frac{1}{m}  - \frac{\eE(R_M|M=m)}{m}.
\label{eq:condexp}
\end{align}
Since $R_M \leq \lambda/n$ so the second term is expected to be negligible with
large sampling rate $n$.  That is, conditional average of $X_i$ is nearly $1/m$
conditioned on $M = m$. In \eqref{eq:exch}, conditioned on $M = m$, the
exchangeable nature of $X_1, X_2, \ldots, X_m$ is used along with $S_m = X_1 +
\ldots + X_m$.

To determine the expectation of average mean-squared
error between $S_1^M$ and equi-spaced grid, consider the conditional expectation
of the following error-term:
\begin{align}
\eE\Bigg[ & \frac{1}{M} \sum_{i = 1}^M \left| S_i - \frac{i}{M} \right|^2 \Bigg|
M = m \Bigg] \\
& = \eE\Bigg[\frac{1}{m} \sum_{i = 1}^m \left\{ \sum_{l = 1}^i \left(X_l -
\frac{1}{m} \right) \right\}^2 \Bigg| M = m \Bigg] \nonumber \\
& = \eE\left[ \frac{1}{m} \sum_{i = 1}^m \sum_{l = 1}^i \sum_{p = 1}^i \left(X_l
- \frac{1}{m} \right) \left(X_p - \frac{1}{m} \right) \Bigg| M = m  \right]
\nonumber \\
& \stackrel{(a)}{=} \eE\Bigg[ \frac{1}{m}  \sum_{i = 1}^m  i  \left(X_1 -
\frac{1}{m} \right)^2 + \nonumber \\
& \mbox{\hspace{1.1cm}} i(i-1) \left(X_1 - \frac{1}{m} \right)  \left(X_2 -
\frac{1}{m} \right) \Bigg| M = m  \Bigg] \\
& = \eE\Bigg[ \frac{(m+1)}{2}  \left(X_1 - \frac{1}{m} \right)^2 + \frac{m^2
-1}{3} \times \nonumber \\
& \mbox{\hspace{2.2cm}} \left(X_1 - \frac{1}{m} \right) \left(X_2 - \frac{1}{m}
\right) \Bigg| M = m  \Bigg] \\
& = \frac{(m+1)}{2} a_m + \frac{m^2 -1}{3} b_m \label{eq:condtermambm}
\end{align}
where $(a)$ follows by exchangeability of $X_1, X_2, \ldots, X_m$ conditioned on
$M = m$, and 
\begin{align}
a_m & := \eE\left[ \left(X_1 - \frac{1}{m} \right)^2 \Bigg| M = m  \right]
\nonumber \\
\mbox{ and } b_m & := \eE\left[ \left(X_1 - \frac{1}{m} \right) \left(X_2 -
\frac{1}{m} \right) \Bigg| M = m  \right] \label{eq:am}
\end{align}
By definition $S_m -1 + R_m = 0$ conditioned on $M = m$, so 
\begin{align}
\eE\left[ (S_m - 1)^2 | M = m \right] = \eE\left[ (R_m)^2 | M = m \right]
\end{align}
and therefore,
\begin{align}
\eE\left[ \left\{ \sum_{i = 1}^m \left(X_i - \frac{1}{m}\right)\right\}^2 \Bigg|
M = m \right] - \eE\left[  R_m^2 \Big| M = m \right] = 0  \nonumber
\end{align}
since $S_m = X_1 + X_2 + \ldots + X_m$. A rearrangement of the above equation
results in,
\begin{align}
m a_m & + m(m-1) b_m = \eE\left[  R_m^2 \Big| M = m \right] \nonumber \\
\mbox{or } b_m & = \frac{1}{m(m-1)} \left( - m a_m + \eE\left[  R_m^2 \Big| M =
m \right] \right). \label{eq:bminam}
\end{align}
With $b_m$ from \eqref{eq:bminam}, the conditional error term in
\eqref{eq:condtermambm} can be rewritten as
\begin{align}
\eE& \Bigg[ \frac{1}{m} \sum_{i = 1}^m \left\{ \sum_{l = 1}^i \left(X_l -
\frac{1}{m} \right) \right\}^2 \Bigg| M = m \Bigg] \\
& = \frac{(m+1)}{2} a_m + \frac{m^2 -1}{3m(m-1)} \left( - m a_m + \eE\left[
R_m^2 \Big| M = m \right] \right) \nonumber \\
& = \frac{(m+1)}{6} a_m + \frac{m + 1}{3m} \eE\left[ R_m^2 \Big| M = m \right]
\nonumber \\
& \leq \frac{(m+1)}{6} a_m + \frac{2}{3} \frac{\lambda^2}{n^2}
\end{align}
where the last step is obtained since $(m+1)/3m \leq 2/3$ and $R_m \leq X_{m+1}
\leq (\lambda/n)$ by assumption on the inter-sample spacing distribution. So,
\begin{align}
\eE \Bigg[ &\frac{1}{M} \sum_{i = 1}^M \left( S_i - \frac{i}{M} \right)^2 \Bigg]
\nonumber \\
& \leq \frac{1}{6} \eE\left( (M+1) a_M \right) + \frac{2}{3}
\frac{\lambda^2}{n^2} \\
& \stackrel{(a)}{=} \frac{1}{6} \eE \left( (M+1) \left(X_1 - \frac{1}{M}
\right)^2 \right) + \frac{2}{3} \frac{\lambda^2}{n^2}\\
& \stackrel{(b)}{\leq} \frac{1}{6} \eE \left( (M+1) \left( 2 X_1^2 +
\frac{2}{M^2}\right) \right) + \frac{2}{3} \frac{\lambda^2}{n^2} \\
& \leq \frac{1}{3} \eE [(M+1) X_1^2]  + \frac{1}{3} \eE \left( \frac{M+1}{M^2}
\right) + \frac{2}{3} \frac{\lambda^2}{n^2} \\
& \stackrel{(c)}{\leq} \frac{1}{3} \eE [(M+1)] \frac{\lambda^2}{n^2} +
\frac{1}{3} \eE \left( \frac{\lambda}{n} + \frac{\lambda^2}{n^2} \right) +
\frac{2}{3} \frac{\lambda^2}{n^2} \\
& \stackrel{(d)}{=} \frac{1}{3} n \frac{\lambda^2}{n^2} + \frac{\lambda}{3n} +
\frac{\lambda^2}{n^2} \leq \frac{C_1}{n}
%
%
\end{align}
for some $C_1 > 0$, which is independent of $n$ and $g(x)$. In the above
inequalities, $(a)$ follows by the definition of $a_m$ in \eqref{eq:am}, $(b)$
follows by $|x_1 - x_2|^2 \leq 2 |x_1|^2 + 2 |x_2|^2$ for any real numbers $x_1$
and $x_2$, $(c)$ follows because $0 < X < \lambda/n$ and therefore $(n/\lambda)
< M < \infty$, and $(d)$ follows using Wald's identity on $(M+1)$ and $\eE(X) =
1/n$~\cite{durrettp1996}. This completes the proof.

\section{Approximation error in integrals by Riemann sum}
\label{ap:riemannapprox}

It is noted that
\begin{align}
|a&[k] - A_{R}[k]| \nonumber \\
& = \bigg| \int_{0}^1 g(x) \exp( - j 2 \pi k x) \mbox{d} x - \nonumber \\
& \mbox{\hspace{2cm}} \frac{1}{M} \sum_{i = 1}^M g \left(\frac{i}{M} \right)
\exp \left( - \frac{j 2 \pi k i}{M} \right) \bigg| \\
& = \bigg| \sum_{i = 1}^M \int_{\frac{i-1}{M}}^{\frac{i}{M}} g(x) \exp( - j 2
\pi k x) \mbox{d}x - \nonumber \\
& \mbox{\hspace{2cm}} \frac{1}{M} \sum_{i = 1}^M g \left(\frac{i}{M} \right)
\exp \left( - \frac{j 2 \pi k i}{M} \right) \bigg| \\
& = \bigg| \sum_{i = 1}^M \frac{1}{M} g(Z_{i,M}) \exp( - j 2 \pi k Z_{i,M}) -
\nonumber \\
& \mbox{\hspace{2cm}} \frac{1}{M} \sum_{i = 1}^M g \left(\frac{i}{M} \right)
\exp \left( - \frac{j 2 \pi k i}{M} \right) \bigg|
\end{align}
where $Z_{i,M} \in (l/M, (l+1)/M)$ is some constant that depends on $g(x) \exp(-
j 2\pi k x)$ by the Lagrange mean-value theorem~\cite{rudinp1976}.  So,
\begin{align}
|a&[k] - A_{R}[k]| \\
& = \frac{1}{M} \bigg| \sum_{i = 1}^M g(Z_{i,M}) \exp( - j 2 \pi k Z_{i,M}) -
\nonumber \\
& \mbox{\hspace{3.5cm}} g \left(\frac{i}{M} \right) \exp \left( - \frac{j2 \pi k
i}{M} \right) \bigg| \\
& \leq \frac{1}{M} \sum_{i = 1}^M \bigg| g(Z_{i,M}) \exp( - j 2 \pi k Z_{i,M}) -
\nonumber \\
& \mbox{\hspace{3.5cm}} g \left(\frac{i}{M} \right) \exp \left( - \frac{j2 \pi k
i}{M} \right) \bigg| \\
& \leq \frac{1}{M} \sum_{i = 1}^M \left| Z_{i,M}  - \frac{i}{M} \right| \left\|
\frac{\mbox{d}}{\mbox{d}x} g(x) \exp(- j 2\pi k x) \right\|_\infty \\
& \leq \frac{1}{M} \sum_{i = 1}^M \frac{1}{M} \left\| \frac{\mbox{d}}{\mbox{d}x}
g(x) \exp(- j 2\pi k x) \right\|_\infty \\
& \leq \frac{1}{M} \left\| \frac{\mbox{d}}{\mbox{d}x} g(x) \exp(- j 2\pi k x)
\right\|_\infty. 
\end{align}
As a result
\begin{align}
\eE \left[ |a[k] - A_{R}[k]|^2 \right] & \leq \eE\left( \frac{C^2}{M^2} \right)
\\
& \leq \frac{C_2^2 \lambda^2}{n^2}.
\end{align}
where $C_2$ is the largest magnitude of the derivative of $g(x) \exp(- j 2\pi
kx)$, and is finite since $g(x)$ and $\exp(- j 2 \pi kx)$ are bounded and
differentiable.

\section{Mean-squared error in the energy estimation}
\label{ap:energyestimate}

Observe that,
\begin{align}
\Bigg| & E_g - \int_{0}^1 g^2(x) \mbox{d}x \Bigg| \\
& = \Bigg| \frac{1}{M} \sum_{i = 1}^M g^2(S_i) + \frac{2}{M}  \sum_{i = 1}^M
g(S_i)W(S_i) + \nonumber \\
& \mbox{\hspace{2.0cm}} \frac{1}{M} \sum_{i = 1}^M \left(W^2(S_i) -
\sigma^2\right)  - \int_{0}^1 g^2(x) \mbox{d}x \Bigg| \\
& \leq \left| \frac{1}{M} \sum_{i = 1}^M g^2(S_i)  - \int_{0}^1 g^2(x) \mbox{d}x
\right| +  \nonumber \\
& \mbox{\hspace{0.5cm}} \left| \frac{2}{M} \sum_{i = 1}^M g(S_i)W(S_i) \right| +
\left| \frac{1}{M}\sum_{i = 1}^M \left(W^2(S_i) - \sigma^2 \right) \right|
\label{eq:threeterms}
\end{align}
The three terms in the above expression will be analyzed one by one and it will
be argued that their mean-squared values do not exceed $O(1/n)$. The first term
can be bounded as
\begin{align}
\Bigg| \frac{1}{M} & \sum_{i = 1}^M g^2(S_i)  - \int_{0}^1 g^2(x) \mbox{d}x
\Bigg| \\
%
%
& \leq \left| \frac{1}{M} \sum_{i = 1}^M g^2 (S_i) - \frac{1}{M} \sum_{i = 1}^M
g^2 \left( \frac{i}{M}\right) \right| + \nonumber \\
& \mbox{\hspace{2.2cm}} \left| \frac{1}{M} \sum_{i = 1}^M g^2 \left(
\frac{i}{M}\right) - \int_{0}^1 g^2(x) \mbox{d}x \right| \\
& \leq \frac{1}{M} \sum_{i = 1}^M \left| g^2 (S_i) - g^2 \left(
\frac{i}{M}\right) \right| + \nonumber \\
& \mbox{\hspace{1.3cm}} \left| \frac{1}{M} \sum_{i = 1}^M g^2 \left(
\frac{i}{M}\right) - \sum_{i = 1}^M \int_{\frac{i-1}{M} }^{\frac{i}{M}} g^2 (x)
\mbox{d}x \right|. \label{eq:g2energy}
\end{align}
By Lagrange mean-value theorem~\cite{rudinp1976} and the continuity of $g(x)$,
it follows that 
\begin{align}
\int_{\frac{i-1}{M}}^{\frac{i}{M}} g^2 (x) \mbox{d}x =  g^2( Z_{i,M} )
\frac{1}{M} \label{eq:lagrangeg2}
\end{align}
where $ \frac{i-1}{M} \leq Z_{i,M} \leq \frac{i}{M}$ or $\left|Z_{i,M} -
\frac{i}{M}\right| \leq \frac{1}{M}$. Further, note that 
\begin{align}
|g^2(x) - g^2(y)| & \leq 2 \left\| g \right\|_\infty |g(x) - g(y)|
\label{eq:g2mvt} \\
& \leq 2 \left\| g \right\|_\infty \left\| g' \right\|_\infty |x-y|.
\label{eq:gmvt}
\end{align}
By the sequential use of \eqref{eq:lagrangeg2}, \eqref{eq:g2mvt}, and
\eqref{eq:gmvt} in \eqref{eq:g2energy}, the following inequalities are obtained:
\begin{align}
\Bigg| \frac{1}{M} & \sum_{i = 1}^M g^2(S_i)  - \int_{0}^1 g^2(x) \mbox{d}x
\Bigg| \\
& \leq \frac{1}{M} \sum_{i = 1}^M \left| g^2(S_i) - g^2 \left( \frac{i}{M}
\right) \right| + \nonumber \\
& \mbox{\hspace{2.4cm}} \frac{1}{M} \sum_{i = 1}^M \left| g^2 \left( \frac{i}{M}
\right) - g^2(Z_{i,M}) \right| \\
& \leq \frac{2 \| g \|_\infty}{M} \sum_{i = 1}^M \left| g(S_i) - g\left(
\frac{i}{M} \right) \right| + \nonumber \\
& \mbox{\hspace{2.7cm}} 2 \|g\| \|g'\| \frac{1}{M} \sum_{i = 1}^M \left|
Z_{i,M} - \frac{i}{M} \right| \\
& \leq \frac{2}{M} \sum_{i = 1}^M \left| g(S_i) - g\left( \frac{i}{M} \right)
\right| +  \frac{4 \pi b}{M}
\end{align}
since $\|g\|_\infty \leq 1$ by assumption and $\|g'\|_\infty \leq 2 \pi b$ from
\eqref{eq:bernstein}. From \eqref{eq:ahatar}, \eqref{eq:ahatarmse}, and
\eqref{eq:epsilonFn}, it can be deduced that
\begin{align}
\eE \left[ \left| \frac{2}{M} \sum_{i = 1}^M \left| g(S_i) - g\left( \frac{i}{M}
\right) \right| \right|^2 \right] \leq \frac{4 \| g' \|^2_\infty C_1}{n}.
\label{eq:onea}
\end{align}
Since $M > (n/\lambda)$, so
\begin{align}
\eE \left( \left| \frac{4\pi b}{M} \right|^2 \right) \leq \frac{16 \pi^2 b^2
\lambda^2}{n^2}. \label{eq:oneb}
\end{align}

The second term in \eqref{eq:threeterms} has a mean-squared error bounded as
follows:
\begin{align}
\eE \Bigg[ \frac{4}{M^2} &  \left| \sum_{i = 1}^M g(S_i)W(S_i) \right|^2 \Bigg]
\nonumber \\
& = \eE \left[ \frac{4}{M^2}  \sum_{i=1}^M \sum_{l=1}^M g(S_i)g(S_l)W(S_i)W(S_l)
\right] \\
& \stackrel{(a)}{=} \eE \left[ \frac{4}{M^2}  \sum_{i = 1}^M g^2(S_i) \sigma^2
\right] \\
& \stackrel{(b)}{=} \eE \left[ \frac{4}{M} \sigma^2 \right] \leq \frac{4
\sigma^2 \lambda}{n} \label{eq:two}
\end{align}
where $(a)$ follows since $W(x)$ process is independent of $M$ and $W(S_i),
W(S_l)$ are independent if $S_i \neq S_l$ by assumption, and $(b)$ follows since
$\|g\|_\infty \leq 1$ and $M > n/\lambda$. 

The third term in \eqref{eq:threeterms} has a mean-squared error bounded as
follows:
\begin{align}
\eE \left[ \left|\frac{1}{M} \sum_{i = 1}^M \left(W^2(S_i) - \sigma^2 \right)
\right|^2 \right] & \stackrel{(a)}{=} \eE \left[ \frac{\sigma^2(W^2)}{M} \right]
\\
& \stackrel{(b)}{\leq} \frac{\sigma^2(W^2)\lambda}{n} \label{eq:three}
\end{align}
where $(a)$ follows since $W(x)$ process is independent of $M$ and $W(S_i),
W(S_j)$ are independent if $S_i \neq S_j$ by assumption and $W^2(x)$ has mean
$\sigma^2$ and variance $\sigma^2(W^2)$ by assumption, and $(b)$ follows since
$M > (n/\lambda)$.

From \eqref{eq:threeterms}, and \eqref{eq:onea}, \eqref{eq:oneb}, \eqref{eq:two}
and \eqref{eq:three}, it follows that
\begin{align}
\eE & \Bigg[ \Big|E_g - \int_{0}^1 g^2(x) \mbox{d}x \Big|^2 \Bigg] \nonumber \\
& \leq  3 \left[ \frac{4 \| g' \|^2_\infty C_1}{n} + \frac{16 \pi^2 b^2
\lambda^2}{n^2} + \frac{4\sigma^2 \lambda}{n} + \frac{\sigma^2(W^2) \lambda}{n}
\right] \nonumber \\ 
& \leq \frac{\tilde{C}}{n} \label{eq:egconv}
\end{align}
where $\tilde{C} > 0$ is some constant independent of $n$ and depends on $g(x)$
only through $b$.

\section*{Acknowledgments} 

The author benefited from discussions with Prof.~Martin Vetterli and
Dr.~Jayakrishnan Unnikrishnan, EPFL, Switzerland, and Prof.~Venkat Anantharam,
EECS, University of California at Berkeley on various topics related to this
paper.

\bibliographystyle{IEEEtran}

\bibliography{../../../../references}

\end{document}